# *ROSAT* PSPC and HRI observations of the composite starburst/Seyfert 2 galaxy NGC 1672

W.N. Brandt,[1] J.P. Halpern[2] and K. Iwasawa[1]

[1] *Institute of Astronomy, Madingley Road, Cambridge CB3 0HA (Internet: wnb@ast.cam.ac.uk, ki@ast.cam.ac.uk)*
[2] *Columbia Astrophysics Laboratory, Columbia University, 538 West 120th Street, New York, NY 10027, USA (Internet: jules@carmen.phys.columbia.edu)*

**ABSTRACT**
The nearby barred spiral galaxy NGC 1672 is thought to have a weak Seyfert nucleus in addition to its strong starburst activity. Observations with the Position Sensitive Proportional Counter (PSPC) and High Resolution Imager (HRI) instruments on board the *ROSAT* X-ray satellite show that three X-ray sources with luminosities (1–2)$\times 10^{40}$ erg s$^{-1}$ are clearly identified with NGC 1672. The strongest X-ray source lies at the nucleus, and the other two lie near the ends of the prominent bar, locations that are also bright in H$\alpha$ and near-infrared images. The nuclear source is resolved by the HRI on about the scale of the recently identified nuclear ring, and one of the sources at the ends of the bar is also probably resolved. The X-ray spectrum of the nuclear source is quite soft, having a Raymond–Smith plasma temperature of $\approx 0.7$ keV and little evidence for intrinsic absorption. The *ROSAT* band X-ray flux of the nuclear source appears to be dominated not by X-ray binary emission but rather by diffuse gas emission. While the properties of the nuclear source are generally supportive of a superbubble interpretation, its large density and emission measure stretch the limits that can be comfortably accommodated by such models. We do not detect direct emission from the putative Seyfert nucleus, although an alternative model for the nuclear source is thermal emission from gas that is photoionized by a hidden Seyfert nucleus. The spectra of the other two X-ray sources are harder than that of the nuclear source, and superbubble models for them have the same strengths and weaknesses.

## 1 INTRODUCTION

### 1.1 Basic facts

NGC 1672 (PKS 0444 − 593) is a $V = 10.1$ barred spiral galaxy of type SB(s)b. It is the principal galaxy in a group that is partially covered by the Large Magellanic Cloud, and a member of the Dorado cloud complex (see fig. 4 of de Vaucouleurs 1975). The radial velocity of NGC 1672 relative to the centroid of the Local Group is 1 140 km s$^{-1}$ (Osmer, Smith & Weedman 1974; hereafter OSW74). It has *four* principal outer arms and these contain many H II regions 2–4 arcsec in size (Sandage & Bedke 1994). On deep optical photographs the arms can be seen to extend out to about 6 arcmin from the centre of NGC 1672, although they become quite faint outside about 3 arcmin from the centre (see Sandage & Bedke 1994). Its bar has a length of 2.4 arcmin and vigorous star formation is seen at its ends (Baumgart & Peterson 1986), as is consistent with calculations of density enhancements associated with bars (see section 9 of Athanassoula 1992 and references therein). Conspicuous dust lanes are present along the leading edges of the bar (Baumgart & Peterson 1986). Elmegreen et al. (1991) suggested that NGC 1672 may have interacted with NGC 1688 and that this interaction may have led to the 'ocular' central shape of NGC 1672. The centres of NGC 1672 and NGC 1688 are separated by 39 arcmin, and NGC 1688 is roughly one third as massive as NGC 1672. We shall adopt a Hubble constant of $H_0 = 50$ km s$^{-1}$ Mpc$^{-1}$ and a cosmological deceleration parameter of $q_0 = \frac{1}{2}$ throughout. This gives a distance to NGC 1672 of 22.8 Mpc (assuming that the peculiar velocity of NGC 1672 is small relative to its Hubble flow velocity). At this distance the bar of NGC 1672 is 16 kpc long and the HRI's spatial resolution ($\approx 5$ arcsec) corresponds to about 550 pc.

### 1.2 The nuclear region

The nucleus of NGC 1672 was noted to be peculiar by Sérsic & Pastoriza (1965). Pastoriza (1973) stated that it has dimensions of $\approx 22 \times 14$ arcsec$^2$ with its longest axis lying in the north-east direction. An isophotal map of the nuclear region can be found in Sérsic (1968). The nuclear spectrum has a polarization of $\leq 1.2$ per cent in the $B$, $R$ and $H$ bands (Brindle et al. 1990). The relative strengths of the optical emission lines in the spectrum taken through a 20-arcsec aperture are similar to those found in ordinary H II regions (OSW74), although sources of gas ionization in addition to normal OB stars cannot be ruled out. Storchi-Bergmann, Wilson & Baldwin (1996; hereafter SWB96) classified the nuclear spectrum as a LINER, although the details of spatially resolved emission-line profiles and ratios in the inner 2 arcsec have also been interpreted as evidence of a composite spectrum of a Seyfert nucleus and H II regions, as described below. NGC 1672 does not have a 'warm' *IRAS* colour in the sense of Sanders et al. (1988).

The manifestations of Seyfert activity in the centre of NGC 1672 are the following.

(1) Optical spectra of the nucleus in a $2 \times 4$ arcsec$^2$ aperture show that its [O III] lines have FWHM $\approx 300$ km s$^{-1}$ while its H$\beta$ line has a FWHM of only 150 km s$^{-1}$ (Véron, Véron & Zuiderwijk 1981). Véron et al. argued that the presence of [O III] broader than H$\beta$ is the signature of a composite Seyfert 2/H II region spectrum, in which the [O III] emission comes primarily from a Seyfert nucleus whose [O III]/H$\beta$ ratio is large. The H$\beta$ flux then comes primarily from a region of circumnuclear starburst activity, where the [O III]/H$\beta$ ratio is small. Díaz (1985) and García-Vargas et al. (1990) presented spatially resolved spectroscopy of NGC 1672 which shows a strong increase in the [O III] line strength near the nucleus. In the very central region ($1.3 \times 1.6$ arcsec) [O III]/H$\beta \approx 1$ while away from this region [O III]/H$\beta$ is significantly less than unity. Their spectra reveal that in the very central region both [O III] and H$\beta$ have about the same FWHM of $\approx 300$ km s$^{-1}$. They also show broad wings of H$\beta$ in absorption, the signature of early-type stellar photospheres. All of these features are present in the nuclear spectra of SWB96 as well, although they classified NGC 1672 as a LINER. We favour slightly the interpretation of Véron et al. (1981), who concluded that the spectrum is a composite Seyfert 2/H II region. The LINER classification may simply be an artefact of applying line-ratio diagnostic diagrams to a two-component spectrum which is spatially unresolved.

(2) Kawara, Nishida & Gregory (1987) claimed to detect a broad Brackett $\gamma$ line of atomic hydrogen from NGC 1672 with a width of $1100 \pm 380$ km s$^{-1}$. In addition, they found that the $K$-band and molecular hydrogen emission of NGC 1672 suggest that it has an active nucleus (see their section III.c). However, Moorwood & Oliva (1988) did not detect the Brackett $\gamma$ line and set an upper limit on it that is a factor of about two below the detection of Kawara et al. (1987).

(3) NGC 1672 has a compact radio source located at its optical nucleus (cf. plate 2 of Harnett 1987). The spectral luminosity of NGC 1672 at 1410 MHz (21 cm) is $2.4 \times 10^{22}$ W Hz$^{-1}$. Comparison with fig. 4 of Ulvestad & Wilson (1984) shows that this is a fairly typical 1410-MHz spectral luminosity for a Seyfert nucleus, although comparison with fig. 1 of Davies (1989) shows that it is not entirely out of the range of 'normal' spirals.

Lindblad & Jörsäter, in preparation, recently used the Australia Telescope National Facility to make synthesis maps of the nuclear region, and they find that it is composed of a small nucleus surrounded by an almost circular ring with a radius of about 5.4 arcsec (cf. section 3 of Sandqvist, Jörsäter & Lindblad 1995). This ring is also seen in H$\alpha$, but the correlation between radio and H$\alpha$ substructure is low. Tovmassian (1968) presented evidence that the 21-cm flux of NGC 1672 increased by at least a factor of 4 between 1962 and 1965 (there are no further data we know of that examine the claimed variability in more detail). The radio spectral index of Harnett (1987) is $0.73 \pm 0.14$, typical of optically thin synchrotron emission.

### 1.3 Previous X-ray observations

NGC 1672 was first detected as an X-ray source by the *Einstein* Observatory (Griffiths, Feigelson & van Speybroeck 1979; Fabbiano, Feigelson & Zamorani 1982; Fabbiano, Kim & Trinchieri 1992). The 4.9-ks *Einstein* observation yielded $129 \pm 16$ counts after background subtraction. The X-ray emission was clearly extended but details of the shape were unclear. Crude spectral fitting to a power-law model was performed by Kruper, Urry & Canizares (1990). The observed 0.2–4.0 keV flux was $\approx 7 \times 10^{-13}$ erg cm$^{-2}$ s$^{-1}$, corresponding to an unabsorbed isotropic luminosity of $\approx 6 \times 10^{40}$ erg s$^{-1}$.

The *Ginga* X-ray satellite made both scanning and pointed observations of NGC 1672 on 1991 August 3 with the Large Area Counter (LAC) instrument (Awaki & Koyama 1993). A hard X-ray source was seen during the scanning observations in the $0.3° \times 4°$ error box. After background subtraction, the *Ginga* LAC count rate during the pointed observation was 2.1 count s$^{-1}$ (the pointed observation had an entrance aperture of $1° \times 2°$ FWHM and no imaging capability within this aperture). The observed 2–10 keV flux was $\approx 3 \times 10^{-12}$ erg cm$^{-2}$ s$^{-1}$, corresponding to a luminosity of $\approx 2 \times 10^{41}$ erg s$^{-1}$. The 2–10 keV photon index was measured to be $1.5 \pm 0.2$ (90 per cent confidence level errors), and the cold column was constrained to be less than $3 \times 10^{22}$ cm$^{-2}$. We shall compare the *Ginga* and *ROSAT* data below.

## 2 OBSERVATIONS, DATA REDUCTION AND ANALYSIS

*ROSAT* PSPC (Trümper 1983; Pfeffermann et al. 1987) observations were made of NGC 1672 starting on 1992 November 29 (RP701021; total raw exposure of 20.0 ks). *ROSAT* HRI observations were made of NGC 1672 starting on 1992 June 24 (RH701022; total raw exposure of 24.4 ks). NGC 1672 was in the centres of the fields of view for both observations. The *ROSAT* observations were performed in the standard 'wobble' mode; to avoid accidental shadowing of sources by the coarse wire grid which forms part of the PSPC entrance window support structure, *ROSAT* performs a slow dithering motion diagonal to the detector axes with a period of $\approx 400$ s and an amplitude of 3 arcmin.

Reduction and analysis of the PSPC and HRI data were performed with the Starlink ASTERIX X-ray data processing system.

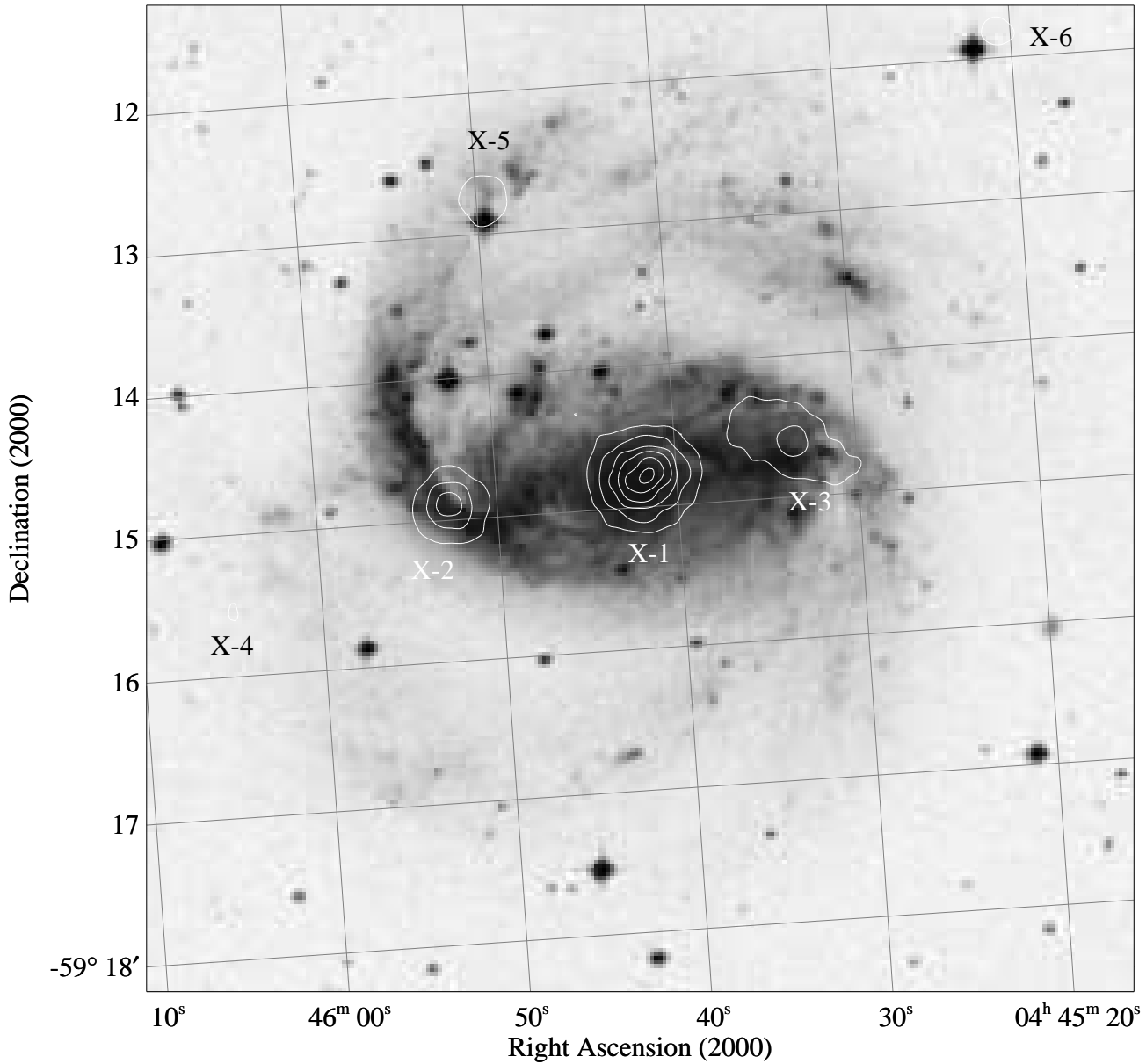

**Figure 1.** Contours of the adaptively smoothed HRI image overlaid on the image of NGC 1672 from the UK Schmidt southern sky survey J plate. Contours are at 5.3, 9.2, 16.0, 27.8, 48.4 and 84.1 per cent of the maximum pixel value (see the text for absolute source fluxes). Note the strong central X-ray source and the locations of X-ray sources near both ends of the bar.

## 2.1 Spatial analysis

### 2.1.1 X-ray sources and naming convention

Fig. 1 shows contours of the adaptively smoothed HRI image overlaid on the image from the UK Schmidt southern sky survey J plate (see section IIb of Lasker et al. 1990 for more information on the optical image). The adaptive smoothing algorithm is described in Rangarajan et al. (1995) and Ebeling, White & Rangarajan (1996). We have set the area that we smooth over by requiring that 25 photons lie within it, and we use a circular tophat smoothing function. Fig. 2 shows the adaptively smoothed PSPC image using PSPC channels 50–200. We have set the area that we smooth over by requiring that 10 photons lie within it. In Table 1 we list the positions of X-ray sources detected near NGC 1672 and give their statistical significances and numbers of counts.

NGC 1688 also lies in the PSPC observation. It is 39 arcmin from the centre of the field of view. As weak sources can be difficult to see when they are far off axis, we have used the ASTERIX point source searching program PSS (Allan 1995) to search the region near NGC 1688 for X-ray sources. We detect an $\approx 5\sigma$ X-ray source (for PSPC channels 50–200) with a most probable position of $\alpha_{2000} = 04^{\rm h}48^{\rm m}24.2^{\rm s}$, $\delta_{2000} = -59°47'54''$ and a positional error of 1.3 arcmin. This position is consistent with that of NGC 1688, although we cannot rule out the possibility that this emission arises from an unrelated background source. After corrections for vignetting and wire shadowing, we derive a count rate in PSPC channels 50–200 of $(5.7 \pm 1.3) \times 10^{-3}$ count s$^{-1}$ (the error bar is for 68 per cent confidence). There are not enough counts in this source for further analysis.

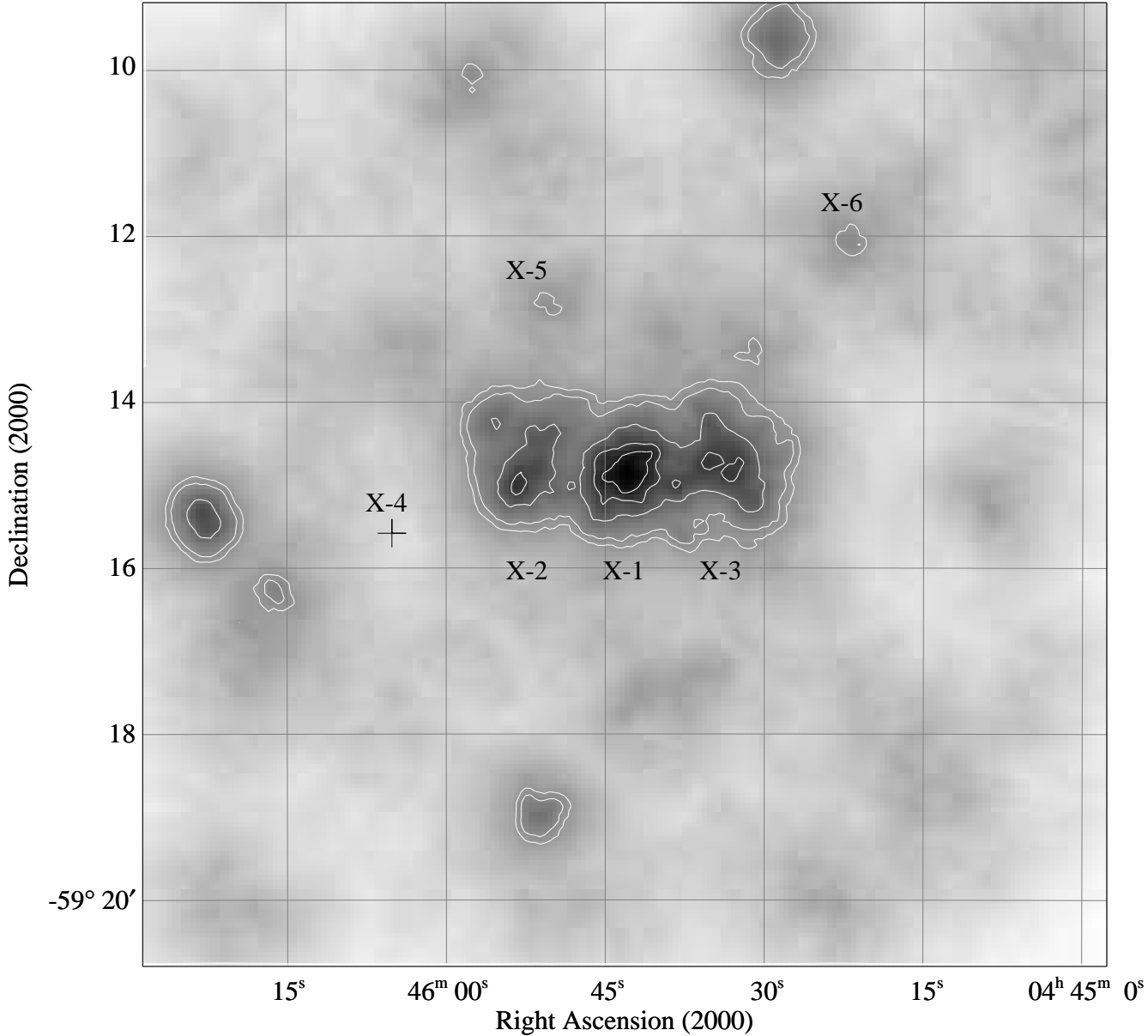

**Figure 2.** The adaptively smoothed PSPC image of NGC 1672 using PSPC channels 50–200. The shading is logarithmic and contours are at 1.1, 1.6, 5.1, 16.0 and 40.3 per cent of the maximum pixel value (see the text for absolute source fluxes).

### 2.1.2  *Association of the X-ray sources with sources at other wavelengths*

The compact radio core of NGC 1672 has a centroid position of $\alpha_{2000} = 04^{\rm h}45^{\rm m}42.48^{\rm s}$, $\delta_{2000} = -59°14'50''$ (Lindblad & Jörsäter, in preparation), and this agrees with the X-ray centroid position of X-1 to within its error. The centroids of X-2 and X-3 are separated by 2.5 arcmin and lie near the ends of the bar. We also note that in the PSPC image

**Table 1.** X-ray sources near to NGC 1672.

| Source name | Source $\alpha_{2000}$ | Source $\delta_{2000}$ | Separation from X-1 | HRI counts | HRI significance | PSPC counts | PSPC significance |
|---|---|---|---|---|---|---|---|
| X-6 | 04 45 21.6 | −59 11 50.9 | 4.0 | $15.5 \pm 6.4$ | 4.1 | $12.6 \pm 4.7$ | 4.5 |
|  | 04 45 28.8 | −59 09 27.3 | 5.7 | $15.5 \pm 5.6$ | 4.1 | $27.8 \pm 6.7$ | 8.3 |
| X-3 | 04 45 33.9 | −59 14 40.3 | 1.1 | $68.9 \pm 13.3$ | 8.0 | $133.6 \pm 11.8$ | 25.1 |
| X-1 | 04 45 42.2 | −59 14 51.1 | 0.0 | $222.8 \pm 17.8$ | 26.4 | $434.6 \pm 21.0$ | 51.8 |
| X-5 | 04 45 49.8 | −59 12 48.9 | 2.3 | $20.0 \pm 6.3$ | 5.4 | $10.6 \pm 4.5$ | 4.1 |
|  | 04 45 51.2 | −59 18 59.9 | 4.3 | $< 14.2$ | — | $17.8 \pm 5.4$ | 5.8 |
| X-2 | 04 45 53.2 | −59 14 57.8 | 1.4 | $48.0 \pm 11.7$ | 10.6 | $113.0 \pm 10.9$ | 19.4 |
| X-4 | 04 46 05.1 | −59 15 34.8 | 3.0 | $17.0 \pm 5.8$ | 4.3 | $< 3.4$ | — |
|  | 04 46 16.5 | −59 16 14.7 | 4.6 | $< 10.1$ | — | $16.3 \pm 5.3$ | 5.3 |
|  | 04 46 23.2 | −59 15 26.8 | 5.3 | $19.4 \pm 6.2$ | 5.3 | $43.0 \pm 8.4$ | 11.7 |

'Source $\alpha_{2000}$' and 'Source $\delta_{2000}$' give the J2000 X-ray centroid positions as determined with the ASTERIX point source searching program PSS (Allan 1995). We quote HRI positions whenever possible but when sources are detected only by the PSPC we quote PSPC positions. HRI positions have errors of $\approx$ 5 arcsec and PSPC positions have errors of $\approx$ 20 arcsec taking into account boresight and other positional errors. The fourth column is the separation in arcmin between the centroid of the source and the centroid of source X-1 (which lies in the centre of NGC 1672).

HRI and PSPC significances are determined using PSS. When determining significances we use the full HRI band and channels 50–200 (corresponding to roughly 0.5–2.0 keV) of the PSPC band. These significances are strictly valid only for point sources (for X-1 and X-3, the two sources that show evidence for spatial extent in the HRI image, there is no question of their reality; see the text). We list all sources near to NGC 1672 that are detected with greater than $4\sigma$ significance by either the HRI or the PSPC.

HRI counts and PSPC counts are the raw numbers of counts after background subtraction. When determining the numbers of counts we use the full HRI band and channels 50–200 of the PSPC band. Count errors are for 68 per cent confidence. For sources other than X-1, X-2 and X-3 we state the numbers of counts determined with PSS (see Allan 1995). For X-1, X-2 and X-3 we have determined the numbers of counts manually. For the HRI we use circular source cells of radii 30, 28 and 30 arcsec for X-1, X-2 and X-3, respectively. We carefully choose these source cells so as to (1) include essentially all of the source counts, (2) retain a reasonably low background contribution and (3) minimize potential cross-source confusion. We use local, circular, source-free cells with radii of 48 arcsec for HRI background subtraction. For the PSPC we use source and background cells as per Table 2. When sources are not detected by one of the two detectors at greater than $4\sigma$, we quote 68 per cent confidence upper limits on their numbers of counts derived using the 'uplim' mode of PSS.

there appears to be some weak emission from a separate pointlike source just to the north-west of X-3. This emission coincides with an optically bright region along one of the arms (compare Figs. 1 and 2).

X-4 is of interest because it lies reasonably close to the centre of NGC 1672 and because it is fairly firmly detected by the HRI yet not by the PSPC (despite the fact that the PSPC observation is deeper). The second fact suggests variability of this source. It lies about 0.5 arcmin off the main part of one of the arms and there is no matching source on the UK Schmidt image or in the NASA Extragalactic Database (NED).

X-5 is located near a bright foreground star which has an optical position of $\alpha_{2000} = 04^{\mathrm{h}}45^{\mathrm{m}}50^{\mathrm{s}}$, $\delta_{2000} = -59°12'56''$. The separation between the optical star and the X-ray centroid is 7 arcsec, and this is the most probable identification for the X-ray source. The star is not listed in SIMBAD and we have not been able to correct its optical position for any proper motion. Similarly, X-6 also appears to be associated with a foreground star.

The other unnamed sources listed in Table 1 do not have any bright optical counterparts. At least some of them are probably background sources.

### 2.1.3 X-ray spatial extents

In Fig. 3 we plot HRI radial brightness profiles of X-1, X-2 and X-3. In making this figure, we have conservatively excluded HRI channels 1–3 to avoid any contamination by UV light (cf. section 3.6 of David et al. 1995). Source X-1 and perhaps source X-3 appear to be extended when compared with the HRI point spread function (PSF; we obtain the PSF from section 3.2.3 of David et al. 1995 and consider the empirical range of PSFs discussed there). We are aware of the effects that aspect solution errors can have on source extensions (cf. Morse 1994), but we are in the fortunate position of having three sources to compare. The fact that X-1 is significantly more extended than X-2 strongly suggests that its apparent extent is real and not an artefact of aspect solution errors. Similarly, the extent of X-3 is also probably real. Examination of the raw HRI image suggests that the extent of X-3 may arise via patchy gas emission or via emission from $\sim$2–3 individual sources, but these data do not allow this possibility to be examined with confidence. None of the sources in the HRI field is bright enough to allow us to perform aspect error correction using the Morse code HRIASPCOR in FTOOLS.

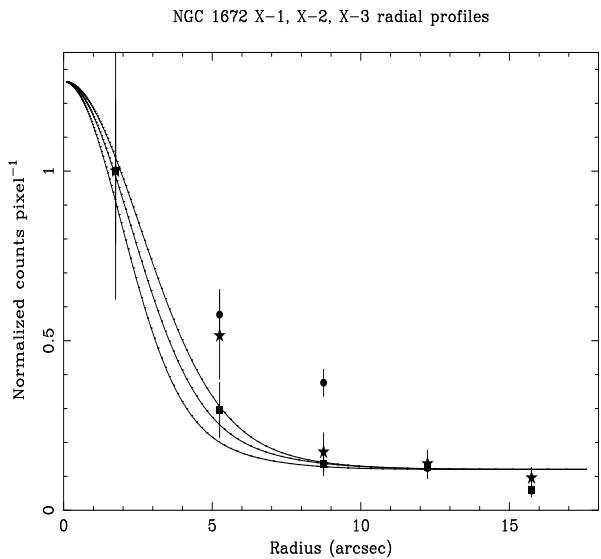

**Figure 3.** HRI radial profiles of NGC 1672 X-1 (circles), X-2 (squares) and X-3 (stars). Each source's first data point is normalized to unity. The expected range of the HRI PSF from David et al. (1995) is graphed as the three solid curves, shifted so that they asymptotically match the background level.

### 2.1.4  *Source count extractions for light curves and spectra*

When preparing the light curves and spectra presented below, we have extracted the source counts from carefully chosen circular source cells that include as many source counts as possible while minimizing cross-source contamination at low energies between X-1, X-2 and X-3. This is difficult in the regions between the sources due to the electronic 'ghost imaging' which widens the point spread function below about 0.3 keV (Hasinger et al. 1992). We discuss this issue in more detail below.

Background counts were subtracted from the source cells using large, nearby circular source-free background cells. Corrections were included for detector dead time, vignetting and shadowing by the coarse mesh window support.

### 2.2  Temporal analysis

Count rates should be averaged over an integer multiple of the 400-s *ROSAT* wobble period when used for source flux determination (cf. Brinkmann et al. 1994). We do not detect any highly statistically significant variability of X-1, X-2 or X-3 within either the PSPC or the HRI data. When we compare count rates for X-1, X-2 and X-3 using the PIMMS software and the spectral models described below, there is no strong evidence for variability between the PSPC and HRI observations.

### 2.3  Spectral analysis

#### 2.3.1  *Relative X-ray energy distributions*

To gain model-independent insight into the spectral differences between X-1, X-2 and X-3, we compared their relative X-ray spectral energy distributions (Fig. 4). Counts from the corrected PSPC source cells were binned so that one data point in Fig. 4 corresponds to 30 PSPC channels, and we have ignored channels below 30 to minimize cross-source contamination. Note that X-1 is the softest of the three sources while X-3 is the hardest.

#### 2.3.2  *Spectral fitting preliminaries and the neutral hydrogen column*

For our spectral fitting, counts from the corrected PSPC source cells were binned into 256-channel, pulse-invariant spectra. We ignored channels 1–8 and rebinned the extracted spectra so that at least 20 source photons were present in each bin. Systematic errors of 2 per cent were added in quadrature to the data point rms errors to account for residual uncertainties in the spectral calibration of the PSPC. We have used the 1993 January response matrix.

We model our X-ray spectra using the spectral models in XSPEC (Shafer et al. 1991). The errors for fit parameters will be quoted for 90 per cent confidence (unless explicitly stated otherwise), taking all free parameters to be of interest other than absolute normalization (Lampton, Margon & Bowyer 1976).

We have used the measurements of Heiles & Cleary (1979) to estimate a Galactic neutral hydrogen column in the direction of NGC 1672. We adopt $N_{\rm H} = 2.0 \times 10^{20}$ cm$^{-2}$ and estimate that the error on this value is about $1 \times 10^{20}$ cm$^{-2}$.

The nuclear region of NGC 1672 has a large apparent Balmer decrement of $H\alpha/H\beta \approx 10$, which OSW74 argue is caused by dust reddening in NGC 1672 with $E(B-V) \approx 1.3$. If we assume a 'Galactic' dust-to-cold-gas ratio, the corresponding neutral hydrogen column density is $N_{\rm H} \approx 7 \times 10^{21}$ cm$^{-2}$ (see section VI of Burstein & Heiles 1978). However, SWB96 argue that the Balmer emission lines need to be corrected for the underlying absorption lines of the early stellar population, which affects $H\beta$ more strongly than $H\alpha$. After subtracting an appropriate stellar template spectrum, the resulting Balmer decrement at the nucleus corresponds to $E(B-V) \approx 0.16$, or $N_{\rm H} \approx 9 \times 10^{20}$ cm$^{-2}$. However, $E(B-V)$ values of 0.4–0.6 are seen within the spatial region corresponding to X-1.

#### 2.3.3  *Spectral fitting to X-1*

The *ROSAT* PSPC spectrum of X-1 is shown in Fig. 5. To avoid confusion by cross-source contamination, we fit only the data above PSPC channel 30 where X-1, X-2 and X-3 are fairly cleanly separated. A power-law model, as might be expected if electron-scattered X-ray emission from a Seyfert nucleus dominates the spectrum, is statistically unacceptable ($\chi^2_\nu = 2.0$) and gives an unphysically steep photon index ($\Gamma > 10$). This is understandable due to the very steep drop-off in X-ray flux above $\approx 1$ keV. If we fit only the data above PSPC channel 50, a simple power-law model can still be ruled out with greater than 95 per cent confidence unless its photon index is greater than 6.0. Such a steep soft X-ray spectrum is never seen in Seyfert galaxies and thus a simple power-law model for X-1 is unphysical. The residuals suggest that a power-law fit is poor because of the robust overall shape of the spectrum and not because of just a few stray

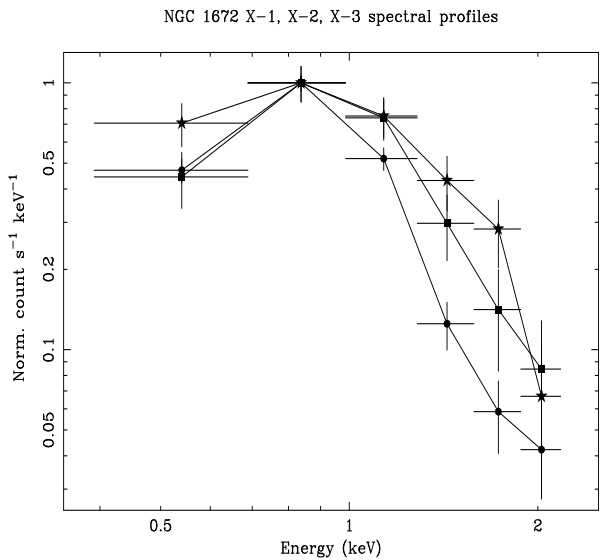

**Figure 4.** PSPC spectra of NGC 1672 X-1 (circles), X-2 (squares) and X-3 (stars). These spectra have been normalized so that their maxima correspond to unity. Note the relative numbers of hard counts from each source.

data points. Deleting sets of points confirms this conclusion.

Electron scattering mirrors in which the atoms in the mirror are not fully stripped of their electrons can also imprint X-ray emission lines on a scattered X-ray spectrum. The strongest such emission lines in the *ROSAT* band are from iron L. We consider an electron scattering mirror that imprints iron L lines by fitting our data to an absorbed power-law model with a Gaussian emission line. The centroid energy of the line is chosen to lie in the range 0.85–1.0 keV, and the line width ($\sigma$) is chosen to lie in the range 0.05–0.20 keV (this is a reasonable model for the iron L complex given the limited spectral resolution of *ROSAT* and the small number of counts in our spectrum). While some choices of line parameters in these ranges give statistically acceptable fits, the underlying photon indices derived from such fits are always significantly larger ($> 2.5$) than are seen in Seyfert 2 galaxies (the soft X-ray spectra of Seyfert 2s are flattened in scattering and have photon indices in the range 1.0–1.6). In addition, the iron L complex equivalent widths derived from our fitting are always larger than 700 eV (and flatter spectra tend to require larger iron L complex equivalent widths). Such equivalent widths, while not impossible, are large compared with what is seen in, for example, Mrk 3 (Iwasawa et al. 1994).

Simple absorbed bremsstrahlung and blackbody models are poor fits to the data and can be rejected with $> 99$ and $> 95$ per cent confidence, respectively. They both leave large systematic residuals in the 0.7–1.0 keV range.

Massive or low-mass X-ray binary sources similar to those in the Milky Way are not likely to make major contributions to X-1's *ROSAT* band flux due to its soft spectrum and fairly steep X-ray drop-off above $\approx 1$ keV (cf. section III.b of Fabbiano 1988; section 4.2 of David, Jones & Forman 1992). They may, of course, contribute significantly at higher energies. Some contribution from 'supersoft' X-ray binaries to the *ROSAT* flux may be possible.

A Raymond–Smith thermal plasma model with cold absorption gives a good fit (see Table 2). The quality of the thermal plasma fit, the arguments above regarding scattered X-ray emission, and the significant spatial extent of X-1 suggest that the starburst activity in NGC 1672 dominates its soft X-ray emission along our line of sight (although we cannot rigorously rule out a substantial contribution of scattered X-rays from a Seyfert nucleus). The derived metal abundance by number relative to the cosmic one (Anders & Grevesse 1989) is $0.12^{+0.40}_{-0.08}$. Such a low derived abundance in the centre of a spiral galaxy appears surprising, and such a low abundance is not suggested by observations at other wavelengths (e.g. SWB96). It must be remembered, however, that due to our limited number of counts we only fitted a single-temperature plasma model to what is probably emission from a multiple-temperature plasma. This simplification can confuse abundance determinations. Fits with two solar-abundance Raymond–Smith plasmas are statistically acceptable and physically reasonable, albeit poorly constrained by these data. In addition, scattered X-rays from the Seyfert nucleus could further complicate the spectrum (although as explained above we suspect they do not dominate it). Fluxes and the isotropic luminosity of X-1 are given in Table 2.

The fitted cold hydrogen column is consistent with the Galactic column, and we do not see evidence for absorption by gas associated with the nuclear $E(B - V)$ of $\gtrsim 0.16$ (SWB96). For reasons discussed in Section 2.3.2, we adopt the reddening analysis of SWB96 rather than that of OSW74. The column associated with this $E(B - V)$ can be statistically accommodated by multi-component model fits, although the best-fitting multi-component models generally have columns of $< 5 \times 10^{20}$ cm$^{-2}$.

If we include the data points below 0.3 keV in our fitting, our results are not significantly changed from those above.

### 2.3.4 Spectral fitting to X-2 and X-3

We shall again use only the data above PSPC channel 30 to reduce cross-source contamination. Due to the small numbers of counts from X-2 and X-3, our spectral models are not tightly constrained, and we shall quote 68.3 per cent confidence errors throughout this section.

A simple absorbed power-law model for X-2's spectrum gives $\Gamma = 2.6^{+2.0}_{-0.8}$, is consistent with the Galactic column, and cannot be ruled out on statistical grounds (it has $\chi^2_\nu = 0.7$). However, given X-2's location and probable starburst nature, we suspect that an absorbed Raymond–Smith thermal plasma model may be more appropriate. Using this model we obtain $N_{\rm H} = (3.9^{+24}_{-3.9}) \times 10^{20}$ cm$^{-2}$, $kT = 1.0^{+5.1}_{-0.7}$ keV and $\chi^2_\nu = 0.9$ (the associated abundance is not well constrained). In the above fitting the parameters are loosely constrained due to the fact that the neutral hydrogen column can become very large. If we assume that there is no significant intrinsic hydrogen column and fix the column at $N_{\rm H} = 2.0 \times 10^{20}$ cm$^{-2}$ (this is at least plausible given the inclination of NGC 1672 and our results for X-1), we obtain the results shown in Table 2.

A simple absorbed power-law model for X-3's spectrum gives $\Gamma = 2.0^{+1.2}_{-0.7}$, is consistent with the Galactic column, and cannot be ruled out on statistical grounds (it has $\chi^2_\nu = $

**Table 2.** Raymond-Smith thermal plasma models of X-ray sources in NGC 1672.

| Quantity | Source X-1 | Source X-2 | Source X-3 |
|---|---|---|---|
| $\alpha_{2000}$ of centre of source cell | 04 45 42 | 04 45 53 | 04 45 33 |
| $\delta_{2000}$ of centre of source cell | $-59\ 14\ 52$ | $-59\ 14\ 45$ | $-59\ 14\ 46$ |
| Radius of source cell (arcmin) | 0.66 | 0.66 | 0.57 |
| $\alpha_{2000}$ of centre of background cell | 04 45 18 | 04 46 07 | 04 45 14 |
| $\delta_{2000}$ of centre of background cell | $-59\ 18\ 26$ | $-59\ 12\ 21$ | $-59\ 17\ 46$ |
| Radius of background cell (arcmin) | 2.20 | 1.81 | 1.86 |
| Raw number of source counts in channels 30–255 | $498 \pm 23$ | $133 \pm 12$ | $154 \pm 13$ |
| $\chi^2_\nu$ | 0.7 | 0.9 | 0.6 |
| $kT$ (keV) | $0.68^{+0.15}_{-0.08}$ | $1.1^{+1.1}_{-0.3}$ | $2.1^{+1.6}_{-1.0}$ |
| Abundance $\star$ | $0.12^{+0.40}_{-0.08}$ | $< 0.48$ | $< 0.78$ |
| $N_{\rm H}/(1 \times 10^{20}\ {\rm cm}^{-2})$ | $1.6^{+2.4}_{-1.6}$ | 2.0† | 2.0† |
| (0.1–2.5 keV Absorbed $F_{\rm X}$)/($1 \times 10^{-13}$ erg cm$^{-2}$ s$^{-1}$) | 2.6 | 0.85 | 1.1 |
| (0.1–2.5 keV Unabsorbed $F_{\rm X}$)/($1 \times 10^{-13}$ erg cm$^{-2}$ s$^{-1}$) | 3.6 | 1.2 | 1.4 |
| (0.1–2.5 keV $L_{\rm X}$)/($1 \times 10^{40}$ erg s$^{-1}$) | 2.1 | 0.69 | 0.83 |
| $EM/(1 \times 10^{63}\ {\rm cm}^{-3})$ | 2.2 | 0.8 | 1.0 |
| $n$ (cm$^{-3}$) | $> 0.17$ | $> 0.30$ | $> 0.17$ |
| $M$ (M$_\odot$) | $< 1.1 \times 10^7$ | $< 2.3 \times 10^6$ | $< 4.8 \times 10^6$ |
| $E$ (erg) | $< 1.4 \times 10^{55}$ | $< 4.8 \times 10^{54}$ | $< 1.9 \times 10^{55}$ |
| $P$ (dyne cm$^{-2}$) | $> 3.8 \times 10^{-10}$ | $> 1.1 \times 10^{-9}$ | $> 1.2 \times 10^{-9}$ |

We quote 90 per cent errors for X-1 and 68 per cent errors for X-2 and X-3. We take all free parameters to be of interest other than absolute normalization. $EM$ is the emission measure, calculated as described in the 'Raymond' model description of Shafer et al. (1991). $n$ is the mean emitting gas density averaged over the source, $M$ is the mass of the emitting gas, $E$ is the thermal energy content of the emitting gas and $P$ is the pressure of the emitting gas. Our calculations of $n$, $M$, $E$ and $P$ parallel those given in section 4.1 of Armus et al. (1995), and we have assumed homogeneous spherical emission regions. We take the radial extent of X-1 to be less than 8 arcsec, which corresponds to 880 pc; the radial extent of X-2 to be less than 4 arcsec, which corresponds to 440 pc; and the radial extent of X-3 to be less than 6 arcsec, which corresponds to 660 pc (see Fig. 3).
$\star$ Note that the *ROSAT* fitted abundances are probably not physically meaningful; see the text.
† Column fixed at this value (see the text).

0.6). As per X-2, however, we suspect that an absorbed Raymond–Smith thermal plasma model may be more appropriate. Using this model we obtain $N_{\rm H} = (2.5^{+11}_{-2.5}) \times 10^{20}$ cm$^{-2}$, $kT = 1.8^{+5.0}_{-0.8}$ keV and $\chi^2_\nu = 0.7$ (the associated abundance is not well constrained). If, as per the previous paragraph, we fix the column at $N_{\rm H} = 2.0 \times 10^{20}$ cm$^{-2}$, we obtain the results shown in Table 2.

### 2.3.5 Comparison of ROSAT and Ginga spectral results

Using the *Ginga* spectral parameters given in Section 1 and the *ROSAT* spectral parameters and data described above, we have compared the spectra from these two satellites. If the *Ginga* spectral data were dominated by a central active nucleus in NGC 1672 and the flux from this nucleus did not vary with time, then one would expect the low-energy end of the *Ginga* spectral model to join fairly smoothly on to the high-energy end of the *ROSAT* PSPC X-1 spectrum. This is not observed. Instead the low-energy end of the *Ginga* spectral model is a factor of $\approx 7$ times higher than the high-energy end of the *ROSAT* PSPC spectrum of X-1, even when the upper limit on the column given by Awaki & Koyama (1993) is used. This suggests that either the *Ginga* spectral data were not dominated by a central active nucleus in NGC 1672 (at least at the low-energy end of the *Ginga* band) or the hard X-ray flux from the central nucleus decreased between the *Ginga* (1991 August 3) and *ROSAT* (1992 November 29) observations. Obvious additional contributors to the *Ginga* spectrum could be X-ray binaries in X-2 and X-3. If we create a *ROSAT* spectrum that is the sum of X-1, X-2 and X-3, its mismatch with the low-energy end of the *Ginga* spectral model is only a factor of about two (even though X-2 and X-3 are weaker overall than X-1, note from Fig. 4 that they are harder). If, as our analysis suggests, X-2 and X-3 contribute to the *Ginga* flux, then the true column density to the nuclear source is probably larger than the upper limit that *Ginga* measured. Of course, sources external to NGC 1672 could also contribute to the *Ginga* flux. Our *ROSAT* field does not have any sources that are much stronger than NGC 1672 in it, but there are four sources of roughly comparable soft X-ray flux that could contribute hard flux.

## 3  DISCUSSION

### 3.1  Source X-1

The soft X-ray isotropic luminosity that we derive for X-1 is only a small fraction of the total isotropic luminosity of the central region. In the near-infrared the $K$-band nuclear isotropic luminosity in a 3 arcsec diameter aperture is $1.0 \times 10^{42}$ erg s$^{-1}$ (Forbes et al. 1992), and significant extended emission is seen as well. In the far-infrared, NGC 1672 has a 40–120 $\mu$m isotropic luminosity of $9.6 \times 10^{43}$ erg s$^{-1}$ (Maia et al. 1994 corrected to $H_0 = 50$ km s$^{-1}$ Mpc$^{-1}$).

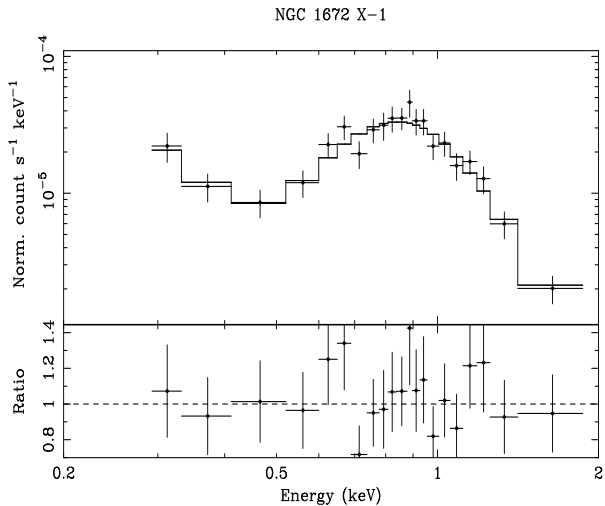

**Figure 5.** PSPC spectrum of NGC 1672 X-1. A Raymond–Smith thermal plasma model with cold absorption is also shown with the corresponding data-to-model ratio.

The 5000-MHz (6-cm) radio luminosity of NGC 1672 is $3.2 \times 10^{38}$ erg s$^{-1}$. Comparison of our spatially resolved X-ray maps and plate 2 of Harnett (1987) is revealing. It shows that, despite the fact that X-1 is stronger than X-2 or X-3 by a factor of only $\approx 2.5$, the central region is at least 15 times more powerful than the regions near X-2 and X-3 at 843 MHz (36 cm). The fact that the central region is $\gtrsim 6$ times more efficient at producing 843-MHz radio flux per unit X-ray flux suggests that either (1) the energy generation mechanism in the central region is different from those near X-2 and X-3 or (2) there is an additional producer of radio flux associated with the central region that does not generate large amounts of observable X-ray flux. As noted before, the 1410-MHz spectral luminosity of NGC 1672 is fairly typical for that of a Seyfert nucleus.

Using a 20-arcsec aperture, OSW74 measured an uncorrected H$\beta$ isotropic luminosity from NGC 1672 of $1.6 \times 10^{40}$ erg s$^{-1}$ (we correct to the distance of Section 1.1 for the OSW74 data here and hereafter). If we adopt an average $E(B-V)$ of $\approx 0.5$ from the starlight-subtracted, near-nuclear measurements of SWB96, then the intrinsic H$\beta$ luminosity is increased to $8 \times 10^{40}$ erg s$^{-1}$. This luminosity requires $1.6 \times 10^{53}$ Lyman continuum photons s$^{-1}$ corresponding to about $2 \times 10^4$ O7 stars. Ward (1988) used the Brackett $\gamma$ data of Kawara et al. (1987) to derive an ionizing photon number of $1.1 \times 10^{53}$ photons s$^{-1}$ (we correct to the distance of Section 1.1). The latter estimate is derived using a smaller aperture size of $6 \times 3.8$ arcsec$^2$.

These estimates for the Lyman continuum photon number above are $\sim 11$ times larger than that from the R136 nebula of 30 Doradus ($\gtrsim 10^{52}$ photons s$^{-1}$; Wang & Helfand 1991), the most luminous H II region in the Local Group. If we assume identical stellar-type distribution functions for the central region of NGC 1672 and the R136 nebula, the number of ionizing stars in the central region of NGC 1672 will be $\sim 11$ times larger than that in the R136 nebula. It is unlikely that O stars directly produce most of X-1's X-rays. For a typical O star X-ray luminosity of $5 \times 10^{33}$ erg s$^{-1}$ (cf. Harnden et al. 1979), about four million O stars would be required. However, most of the X-rays associated with O stars do not come directly from the stars themselves but come instead from the energy that their winds and supernovae deposit in the interstellar medium. If we scale the soft X-ray luminosity of the R136 nebula ($(1$–$4) \times 10^{37}$ erg s$^{-1}$; Wang & Helfand 1991) by the factor of 11 mentioned above, we derive a characteristic soft X-ray luminosity of only $(1$–$4.4) \times 10^{38}$ erg s$^{-1}$. This is to be compared with X-1's isotropic luminosity in soft X-rays of $2 \times 10^{40}$ erg s$^{-1}$. Hence the nuclear activity appears to be significantly more efficient in generating X-rays per Lyman continuum photon than the activity in R136, perhaps due to a larger fractional contribution from non-stellar processes such as supernovae, or heating by a hidden Seyfert nucleus.

In Table 2 we present emission measures for our thermal plasma fits as well as quantities that we estimate from these emission measures. The internal pressure of X-1 corresponds to 140 times that of the nearby hot interstellar medium (Bowyer et al. 1995). The thermal plasma models for X-1, X-2 and X-3 imply surprisingly large densities and thermal energy contents. Because the apparent radius of each source is relatively small, we are forced to consider particle densities of order 0.2 cm$^{-3}$ and larger. X-1, X-2 and X-3 contain energies corresponding to about $10^3$–$10^4$ supernovae. Although the sizes and temperatures of these regions are comparable to those expected for superbubbles blown by this many supernovae over a $10^7$-yr period, their inferred internal densities are $\sim 50$ times greater, even when the evaporation of the cold dense shell to the interior is included (equation 5 of MacLow & McCray 1988; Armus et al. 1995; Heckman et al. 1995). Note that the densities we derive for our *hot* superbubble gas are comparable to those seen in the *cold* interstellar medium in the regions between diffuse and molecular clouds (see table 11.1 of Spitzer 1978). What aspects of this scenario could be altered to retain a plausible thermal plasma model? First, the derived emission measures might be overestimates if the fitted abundances are in error. It is difficult, if not impossible, to measure starburst abundances reliably using *ROSAT* data. If the actual metal abundances were the solar ones, for example, then we estimate with XSPEC that the derived emission measure of X-1 could be lower by a factor of about 6. SWB96 derived abundances that are slightly higher than solar in the nuclear region of NGC 1672, and approximately solar along the bar. However, since the average number density scales only as the square root of the emission measure it is hard to imagine that the density could be reduced by more than a factor of four with such a correction. Secondly, if the hot medium were clumped as a result of the supernova explosions taking place among a large number of giant molecular clouds which have been completely shocked, but have not finished expanding, then the required mass and energy content would be reduced since these quantities scale as the filling factor to the one-half power (see section 4.1 of Armus et al. 1995; the filling factor is less than unity). Together, these two effects may render plausible a model in which one or more of the main X-ray sources in NGC 1672 are extremely luminous ex-amples of superbubbles of the type envisioned by MacLow & McCray (1988), and thought to be seen by *ROSAT* in NGC 5408 (Fabian & Ward 1993), NGC 2146 (Armus et al. 1995) and NGC 1569 (Heckman et al. 1995). A superbubble may also account for the small X-ray absorption measured for X-1, since it could deposit hot gas above the disc. The spatial extents of X-1 and X-3 appear to be larger than the $\sim$ 150-pc scale height of a typical galactic interstellar medium, and this fact also suggests that a significant amount of X-ray emitting gas may lie above the starburst region (which is probably confined to about the height of the molecular cloud layer).

Alternatively, for the nuclear source X-1 we may consider a contribution due to photoionization heating by a hidden Seyfert nucleus. Adopting the fitted thermal plasma model as a first approximation, the emission measure is considerably smaller than that of extended X-ray sources in other Seyfert galaxies. For example, the X-ray emission extended over the inner 6-kpc radius in NGC 1068 requires an emission measure $\sim 7 \times 10^{64}$ cm$^{-3}$ and an electron density $n \sim 0.15$ at an assumed temperature of $10^7$ K (Halpern 1992). This medium would be in rough pressure balance with the 'diffuse ionized medium' seen in [N II] and H$\alpha$ (Bland-Hawthorn, Sokolowski & Cecil 1991). Similarly, Wilson et al. (1992) argued that $(4–9) \times 10^7$ M$_\odot$ of hot gas would be required to account for the extended *nuclear* X-rays in NGC 1068. In NGC 4151 (Morse et al. 1995), the extended X-ray emission has a luminosity of $\sim 1.7 \times 10^{41}$ erg s$^{-1}$, which would require a medium of $T \sim 10^7$ K and $n \sim 0.3$ if it were to be in pressure equilibrium with the extended narrow-line-region clouds. Both NGC 1068 and NGC 4151 have luminous enough ionizing continua to heat their gas to $10^7$ K and render it optically thin to 1-keV X-rays. The required ionization parameter $\xi = L/nr^2$ is $\sim 100$ for a $\Gamma = 2$ power law. For the case of NGC 1672, taking $n = 0.17$ and $r = 880$ pc requires an ionizing luminosity $L \sim 1 \times 10^{44}$ erg s$^{-1}$. This would be a fairly strong Seyfert, and its ionizing luminosity would be comparable to its far-infrared luminosity, although its *IRAS* colours show no evidence for a luminous Seyfert component.

The required intrinsic X-ray luminosity in the Seyfert scenario is about 500 times larger than the 2–10 keV luminosity observed by *Ginga*, and suggests that, if a Seyfert is photoionizing the central region, we are seeing little if any scattered X-ray flux from the nucleus. The properties of X-1 listed in Table 2 allow it to have a Thomson depth of only $\sim 3 \times 10^{-4}$, which is not great enough to account for the flux observed by *Ginga*, although it is roughly compatible with the *soft* X-ray flux being scattered. For either scattering or photoionization heating to account for X-1, the geometry would have to be such as to hide the broad-line region (if any) and the continuum source, but with a relatively large opening angle so as to cause little azimuthal asymmetry in the extended X-ray source, and little reprocessed warm infrared emission. The existence of Seyfert 2 galaxies whose observed X-ray luminosity is even less than $10^{40}$ erg s$^{-1}$ can perhaps be similarly explained by the lack of a suitable scattering mirror.

### 3.2 Sources X-2 and X-3

Sources X-2 and X-3 have not been the subjects of detailed studies in the past. Baumgart & Peterson (1986) commented upon vigorous star formation at the ends of the bar seen in near-infrared photographic plates, but did not present quantitative information. Similarly, H$\alpha$ hotspots at the ends of the bar, as well as along two of the four arms, can be seen in fig. 3 of Sérsic & Calderón (1979), and fig. 2(c) of SWB96. The abundances seem to be about solar in these regions, and the extinction is $E(B − V) \approx 0.27$ for both (SWB96). X-2 and X-3 are each over 150 times more luminous in soft X-rays than the R136 nebula of 30 Doradus. Their emission can be modelled by thermal gas emission, although due to their faintness we cannot rigorously rule out a substantial contribution from X-ray binaries. An interpretation in terms of superbubbles follows the same analysis as for the nuclear source and encounters similarly extreme densities. A significant X-ray binary contribution would help to reduce these densities.

### ACKNOWLEDGMENTS

We acknowledge discussions with D. Calzetti, H. Ebeling, A.C. Fabian, S. Sigurdsson, E. Terlevich, R. Terlevich, M. Ward and the members of the Institute of Astronomy X-ray group. We thank P.O. Lindblad and S. Jörsäter for kindly sharing some of their results prior to publication. We thank an anonymous referee for helpful comments. We acknowledge financial support from the United States National Science Foundation and the British Overseas Research Studentship Programme (WNB), NASA grant NAG 5-1935 (JPH) and the JSPC and the British Council (KI). The *ROSAT* project is supported by the Bundesministerium für Forschung und Technologie (BMFT) and the Max-Planck society. Much of our analysis has relied on the ASTERIX and FTOOLS X-ray data processing systems and the XSPEC X-ray spectral fitting software, and we thank the people who have created and maintain this software. This research has made use of the SIMBAD database, operated at CDS, Strasbourg, France, and the NASA/IPAC extragalactic database (Helou et al. 1991) which is operated by the Jet Propulsion Laboratory, Caltech.

This paper has been produced using the Blackwell Scientific Publications T$_{\!E}$X macros.